%% file: main.tex
\newcommand{\CLASSINPUTtoptextmargin}{1.91cm}
\newcommand{\CLASSINPUTbottomtextmargin}{1.91cm}
\documentclass[conference]{IEEEtran}
\IEEEoverridecommandlockouts

\usepackage{cite, url}
\usepackage[dvipsnames]{xcolor}
\usepackage{soul}
\usepackage{colortbl}
\usepackage[T1]{fontenc}

\usepackage{tikz, forest}
\usepackage{etoolbox}
\usepackage{epstopdf}

\ifCLASSINFOpdf
\usepackage{graphicx}
\usepackage{tabularx}
\usepackage{amssymb}
\usepackage{float}
\usepackage{adjustbox}
\let\labelindent\relax
\usepackage{enumitem}
\graphicspath{{./pdf/}{./jpeg/}}
\else
\fi

\usepackage{algorithm}

\usepackage[noend]{algpseudocode}

\makeatletter

\newsavebox{\measure@tikzpicture}
\NewEnviron{scaletikzpicturetowidth}[1]{%
  \def\tikz@width{#1}%
  \begin{lrbox}{\measure@tikzpicture}%
  \BODY
  \end{lrbox}%
  \pgfmathparse{#1/\wd\measure@tikzpicture}%
  \BODY
}
\renewcommand{\ALG@beginalgorithmic}{\footnotesize}
\def\BState{\State\hskip-\ALG@thistlm}
\makeatother

\usetikzlibrary{arrows.meta}
\usetikzlibrary{backgrounds}
\usetikzlibrary{matrix}

\ifCLASSOPTIONcompsoc
  \usepackage[caption=false,font=normalsize,labelfont=sf,textfont=sf]{subfig}
\else
  \usepackage[caption=false,font=footnotesize]{subfig}
\fi

\usepackage{array}
\usepackage{pbox}

\usepackage[cmex10]{amsmath}
\usetikzlibrary{calc}
\usetikzlibrary{positioning}
\usetikzlibrary{decorations.text}
\usetikzlibrary{arrows.meta}
\usepackage{multirow}

\tikzstyle myBG=[line width=3pt,opacity=1]

\usepackage[absolute,showboxes]{textpos}

\TPMargin{5pt}

\begin{document}
%

\title{P4BFT: Hardware-Accelerated \\ Byzantine-Resilient Network Control Plane}

\author{
	\IEEEauthorblockN{Ermin Sakic\IEEEauthorrefmark{1}\IEEEauthorrefmark{2}, Nemanja Deric\IEEEauthorrefmark{1}, Endri Goshi\IEEEauthorrefmark{1}, Wolfgang Kellerer\IEEEauthorrefmark{1}}%
\IEEEauthorrefmark{1}Technical University Munich, Germany, 
\IEEEauthorrefmark{2} Siemens AG, Germany \\
\footnotesize{E-Mail:\IEEEauthorrefmark{1}\{ermin.sakic, nemanja.deric, endri.goshi, wolfgang.kellerer\}@tum.de, \IEEEauthorrefmark{2} ermin.sakic@siemens.com}
}

\maketitle
\begin{abstract}
	Byzantine Fault Tolerance (BFT) enables correct operation of distributed, i.e., replicated applications in the face of malicious take-over and faulty/buggy individual instances. Recently, BFT designs have gained traction in the context of Software Defined Networking (SDN). In SDN, controller replicas are distributed and their state replicated for high availability purposes. Malicious controller replicas, however, may destabilize the control plane and manipulate the data plane, thus motivating the BFT requirement. Nonetheless, deploying BFT in practice comes at a disadvantage of increased traffic load stemming from replicated controllers, as well as a requirement for proprietary switch functionalities, thus putting strain on switches' control plane where particular BFT actions must be executed in software. 

	P4BFT leverages an optimal strategy to decrease the total amount of messages transmitted to switches that are the configuration targets of SDN controllers. It does so by means of message comparison and deduction of correct messages in the determined optimal locations in the data plane. In terms of the incurred control plane load, our P4-based data plane extensions outperform the existing solutions by $\sim33.2$\% and $\sim40.2$\% on average, in random $128$-switch and Fat-Tree/Internet2 topologies, respectively. To validate the correctness and performance gains of P4BFT, we deploy \emph{bmv2} and \emph{Netronome Agilio SmartNIC}-based topologies. The advantages of P4BFT can thus be reproduced both with software switches and "commodity" P4-enabled hardware. A hardware-accelerated controller packet comparison procedure results in an average $96.4$ \% decrease in processing delay per request compared to existing software approaches. 
\end{abstract}


%
\IEEEpeerreviewmaketitle

\input{introduction.tex}

\input{relatedwork.tex}
\input{models.tex}

\input{discussion.tex}
\input{conclusion.tex}

\section*{Acknowledgment}

This work has received funding from European Commission's H2020 research and innovation programme under grant agreement no. 780315 SEMIoTICS and from the German Research Foundation (DFG) under the grant number KE 1863/8-1. We are grateful to Cristian Bermudez Serna, Dr. Johannes Riedl and the anonymous reviewers for their useful feedback and comments.


%



\bibliographystyle{IEEEtran}
\bibliography{IEEEabrv,qos}

\end{document}

%% file: introduction.tex
\section{Introduction}
State-of-the-art failure-tolerant SDN controllers base their state distribution on crash-tolerant consensus approaches. Such approaches comprise single-leader operation, where leader replica decides on the ordering of client updates. After confirming the update with the follower majority, the leader triggers the cluster-wide commit operation and acknowledges the update with the requesting client. RAFT algorithm \cite{howard2015raft} realizes this approach, and is implemented in OpenDaylight \cite{odl} and ONOS \cite{berde2014onos}. RAFT is, however, unable to distinguish malicious / incorrect from correct controller decisions, and can easily be manipulated by an adversary in possession of the leader replica \cite{bftraft}. Recently, Byzantine Fault Tolerance (BFT)-enabled controllers were proposed for the purpose of enabling correct consensus in scenarios where a \emph{subset of controllers is faulty due to a malicious adversary or internal bugs} \cite{li2014byzantine, mohan2017primary, morph}. In BFT-enabled SDN, multiple controllers act as replicated state machines and hence process incoming client requests individually. Thus with BFT, each controller of a single administrative domain transmits an output of their computation to the target switch. The outputs of controllers are then collected by trusted configuration targets (e.g., switches) and compared for payload matching for the purpose of correct message identification. 

In \emph{in-band} \cite{schiff2016band} deployments, where application flows share the same infrastructure as the control flows, the traffic arriving from controller replicas imposes a non-negligible overhead \cite{muqaddas2016inter}. Similarly, comparing and processing controller messages in the switches' software-based control plane causes additional delays and CPU load \cite{morph}, leading to longer reconfigurations. Moreover, the comparison of control packets is implemented as a proprietary non-standardized switch function, thus unsupported in off-the-shelf devices.

In this work, we investigate the benefits of offloading the procedure of comparison of controller outputs, required for correct BFT operation, to carefully selected network switches. 
By minimizing the distance between the processing nodes and controller clusters / individual controller instances, we decrease the network load imposed by BFT operation. P4BFT's P4-enabled pipeline is in charge of controller packet collection, correct packet identification and its forwarding to the destination nodes, thus minimizing accesses to the switches' software control plane and effectively outperforming the existing software-based solutions.

\section{Background and Problem Statement}
\label{background}

BFT has recently been investigated in the context of distributed SDN control plane \cite{li2014byzantine, mohan2017primary, morph, iccbft}. In \cite{li2014byzantine, mohan2017primary}, $3F_M+1$ controller replicas are required to tolerate up to $F_M$ Byzantine failures. MORPH \cite{morph} requires $2F_M+F_A+1$ replicas in order to tolerate up to $F_M$ Byzantine and $F_A$ availability-induced failures. The presented models assume the deployment of SDN controllers as a set of replicated state machines, where clients submit inputs to the controllers, that process them in isolation and subsequently send the computed outputs to the target destination (i.e., reconfiguration messages to destination switches). They assume trusted platform execution and a mechanism in the destination switch, capable of comparison of the controller messages and deduction of the correct message. Namely, after receiving $F_M+1$ matching payloads, the observed message is regarded as correct and the containing configuration is applied. 

The presented models are sub-optimal in a few regards. First, they assume the collection and processing of controller messages exclusively in the receiver nodes (configuration targets). Propagation of each controller message can carry a large system footprint in large-scale in-band controlled networks, thus imposing a non-negligible load on the data plane. Second, neither of the models detail the overhead of message comparison procedure in the target switches. The realizations presented in \cite{li2014byzantine, mohan2017primary, morph, iccbft} realize the packet comparison procedure solely in software. The non-deterministic / varied latency imposed by the software switching may, however, be limiting in specific use cases, such as in the failure scenarios in critical infrastructure networks \cite{sakicresponse} or in 5G scenarios \cite{5g}. This motivates a hardware-accelerated BFT design that minimizes the processing delays. 

\subsection{Our contribution}
We introduce and experimentally validate the P4BFT design, which builds upon \cite{li2014byzantine, mohan2017primary, morph} and adds further optimizations:
\begin{itemize}
	\item It allows for collection of controllers' packets and their comparison in \emph{processing} nodes, as well as for relaying of deduced correct packets to the destinations;
	\item It selects the optimal processing nodes at per-destination-switch granularity. The proposed objective minimizes the control plane load and reconfiguration time, while considering constraints related to the switches' processing capacity and the upper-bound reconfiguration delay;
	\item It executes in software, e.g., in P4 switch behavioral model (\emph{bmv2}\footnote{P4 Software Switch - \url{https://github.com/p4lang/behavioral-model}}), or in a physical, e.g., Netronome SmartNIC\footnote{Netronome Agilio\textregistered CX 2x10GbE SmartNIC Product Brief - \url{https://www.netronome.com/media/documents/PB_Agilio_CX_2x10GbE.pdf}} environment. Correctness, processing time and deployment flexibility are validated in both platforms.
\end{itemize}

We present the evaluation results of P4BFT for well-known and randomized network topologies and varied controller and cluster sizes and their placements. To the best of our knowledge, this is the first implementation of a BFT-enabled solution on a hardware platform, allowing for accelerated packet processing and low-latency malicious controller detection time.

\textbf{Paper Structure}: Related work is presented in Section \ref{relatedwork}. Section \ref{model} details the P4BFT co-design of the control and data plane as well as the optimization procedure. Section \ref{discussion} presents the evaluation methodology and discusses the empirically measured performance of P4BFT in software- and hardware-based data planes. Section \ref{conclusion} concludes this paper.

%% file: relatedwork.tex
\section{Related Work}
\label{relatedwork}

\subsubsection{BFT variations in SDN context} In the context of centralized network control, BFT is still a relatively novel area of research. Reference solutions \cite{li2014byzantine, mohan2017primary, morph}, assume the comparison of configuration messages, transmitted by the controller replicas, in the switch destined as the configuration target. With P4BFT, we investigate the flexibility advantages of message processing in any node capable of message collection and processing, thus allowing for a footprint minimization. \cite{mohan2017primary} and \cite{morph} discuss the strategy for minimization of no. of matching messages required to deduce correct controller decisions, which we adopt in this work as well. \cite{iccbft} discusses the benefit of disaggregation of BFT consensus groups in the SDN control plane into multiple controller cluster partitions, thus enabling higher scalability than possible with \cite{mohan2017primary} and \cite{morph}. While compatible with \cite{iccbft}, our work focuses on scalability enhancements and footprint minimization by means of data-plane reconfiguration for realizing more efficient packet comparison.

\subsubsection{Data Plane-accelerated Service Execution} Recently, Dang et al. \cite{dang2016paxos} have portrayed the benefits of offloading coordination services for reaching consensus to the data plane, on the example of a Paxos implementation in P4 language. In this paper, we investigate if a similar claim can be transferred to BFT algorithms in SDN context. In the same spirit, in \cite{raftp4}, end-hosts partially offload the log replication and log commitment operations of RAFT consensus algorithm to neighboring P4 devices, thus accelerating the overall commit time. In the context of in-network computation, Sapio et al. \cite{sapio2017network} discuss the benefit of data aggregation offloading to constrained network devices for the purpose of data reduction and minimization of workers' computation time.

%% file: models.tex
\newcolumntype{P}[1]{>{\centering\arraybackslash}p{#1}}
\newcolumntype{M}[1]{>{\centering\arraybackslash}m{#1}}

\section{System Model and Design}
\label{model}
\subsection{P4BFT System Model}
We consider a typical SDN architecture allowing for flexible function execution on the networking switches for the purpose of BFT system operation. The flexibility of in-network function execution is bounded by the limitation of the data plane programming interface (i.e., the $\text{P4}_{16}$ \cite{bosshart2014p4} language specification in the case of P4BFT). The control plane communication between the switches and controllers and in-between the controllers is realized using an in-band control channel \cite{schiff2016band}. In order to prevent faulty replicas from impersonating correct replicas, controllers authenticate each message using Message Authentication Codes (assuming pre-shared symmetric keys for each pair) \cite{eischer2017}. Similarly, switches that are in charge of message comparison and message propagation to the configuration targets must be capable of signature generation using the processed payload and their secret key. 


In P4BFT, controllers calculate their decisions in isolation from each other, and transmit them to the destination switch. Control packets are intercepted by the \emph{processing} nodes (i.e., processing switches) responsible for decisions destined for the target switch. In order to collect and compare control packets, we assume packet header fields that include the \texttt{client\_request\_id}, \texttt{controller\_id}, \texttt{destination\_switch\_id} (e.g., MAC/IP address), the \texttt{payload} (controller-decided configuration) and the optional \texttt{signature} field (denoting if a packet has already been processed by a processing node). Clients must include the \texttt{client\_request\_id} field in their controller requests.

Apart from distinguishing correct from malicious/incorrect messages, P4BFT allows for identification and exclusion of \emph{faulty} controller replicas. P4BFT's architectural model assumes three entities, each with a distinguished role:

\textbf{1) Network controllers} enforce forwarding plane configurations based on internal decision making. For simplification, each controller replica of an administrative domain serves each client request. Each \emph{correct} replica maintains internal state information (e.g., resource reservations) matching to that of other \emph{correct} instances. In the case of a controller with diverged state, i.e., as a result of corrupted operation or a malicious adversary take-over, the \emph{incorrect} controllers' computation outputs may differentiate from the correct ones. 

\textbf{2) P4-enabled switches} forward the control and application packets. Depending on the output of Reassigner's optimization step, a switch may be assigned the \emph{processing} node role, i.e., become in charge of comparing outputs computed by different controllers, destined for \emph{itself} or \emph{other} configuration targets. A processing node compares messages sent out by different controllers and distinguishes the correct ones. On identification of a faulty controller, it declares the faulty replica to the Reassigner. In contrast to \cite{li2014byzantine, mohan2017primary, morph}, P4BFT enables control packet comparison for packets destined for remote targets. 

\textbf{3) Reassigner} is responsible for two tasks: 

\emph{Task 1}: 
It dynamically reassigns the controller-switch connections based on the events collected from the detection mechanism of the switches, i.e., upon their detection, it excludes faulty controllers from the assignment procedure. It furthermore ensures that a minimum number of required controllers, necessary to tolerate a number of availability failures $F_A$ and malicious failures $F_M$, are loaded and associated with each switch. This task is also discussed in \cite{morph, mohan2017primary}.

\emph{Task 2}: It maps a \emph{processing} node, in charge of controller messages' comparison, to each destination switch. Based on the result of this optimization, switches gain the responsibility of control packets processing. The output of the optimization procedure is the \emph{Processing Table}, necessary to identify the switches responsible for comparison of controller messages. Additionally, the Reassigner computes the \emph{Forwarding Tables}, necessary for forwarding of controller messages to processing nodes and reconfiguration targets. Given the no. of controllers and the user-configurable parameter of max. tolerated Byzantine failures $F_M$, Reassigner reports to processing nodes the no. of \emph{necessary matching messages} that must be collected prior to marking a controller message as correct. 

\subsection{Finding the Optimal Processing Nodes}
\label{optimization}
\begin{figure*}
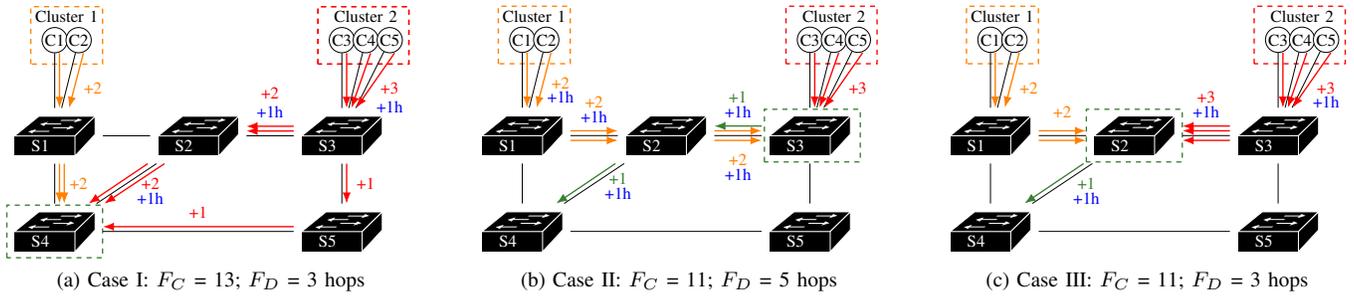

	\centering
	\include{optimization}
	\caption{For brevity we depict the control flows destined only for configuration target \texttt{S4}. The orange and red blocks represent an exemplary cluster separation of $5$ controllers into groups of $2$ and $3$ controllers, respectively. The green dashed block highlights the processing node responsible for comparing the controller messages destined for \texttt{S4}. Figure (a) presents the unoptimized case as per \cite{li2014byzantine, morph, mohan2017primary}, where \texttt{S4} collects and processes control messages destined for itself, thus resulting in a control plane load of $F_C=13$ and a delay on critical path (marked with blue labels) of $F_D=3$ hops (assuming edge weights of $1$). By optimizing for the total communication overhead, the total $F_C$ can be decreased to $11$, as portrayed in Figure (b). Contrary to (a), in (b) processing of packets destined for \texttt{S4} is offloaded to the processing node \texttt{S3}. However, additional delay is incurred by the traversal of path \texttt{S1}-\texttt{S2}-\texttt{S3}-\texttt{S2}-\texttt{S4} for the control messages sourced in Cluster $1$. Multi-objective optimization according to P4BFT, that aims to minimize both the communication overhead and control plane delay instead selects \texttt{S2} as the optimal processing node (ref. Figure (c)), thus minimizing both $F_C$ and $F_D$. } 
	\label{fig:failure_model}
\end{figure*}

The optimization methodology allows for minimization of the experienced switch reconfiguration delay, as well as the decrease of the total network load introduced by the exchanged controller packets. When a switch is assigned the \emph{processing} node role for itself or another target switch, it collects the control packets destined for the target switch and deduces the correct payload on-the-fly, it next forwards a single packet copy containing the correct controller message to the destination switch. Consider Fig. \ref{fig:failure_model}a). If control packet comparison is done only at the target switch (as in prior works), a request for \texttt{S4} creates a total footprint of $F_C=13$ packets in the data plane (the sum of Cluster 1 and Cluster 2 utilizations of $4$ and $9$, respectively). In contrast, if the processing is executed in \texttt{S3} (as depicted in Fig. \ref{fig:failure_model}b)), the total experienced footprint can be decreased to $F_C=11$. Therefore, in order to minimize the total control plane footprint, we identify an optimal processing node for each target switch, based on a given topology, placement of controllers and the processing nodes' capacity constraints. If we additionally extend the optimization to a multi-objective formulation by considering the delay metric, the total traversed critical path between the controller furthest away from the configuration target would equal $F_D=3$ in the worst case (ref. Fig. \ref{fig:failure_model}c)), i.e., $3$  hops assuming a delay weight of $1$ per hop. Additionally, this assignment also has the minimized communication overhead of $F_C=11$.

\definecolor{Gray}{gray}{0.9}
\begin{table}[H]
	\centering
	\caption{Parameters used in the model}	
	\resizebox{\columnwidth}{!}{
		\begin{tabular}{|M{3.7cm}|M{7cm}|}
			\hline
			\textbf{Symbol} & \textbf{Description}\\
			\hline
			\rowcolor{Gray}
			$\mathcal{V}:\{S_1, S_2, ..., S_n\}, n\in \mathbb{Z}^+$ & Set of all switch nodes in the topology.\\
			$\mathcal{C}:\{C_1, C_2, ..., C_n\}, n\in \mathbb{Z}^+$ & Set of all controllers connected to the topology.\\
			\rowcolor{Gray}
			$\mathcal{D}:\{d_{i,j,k}, \forall i,j,k \in \mathcal{V}\}$ & Set of delay values for path from $i$ to $k$, passing through $j$.\\
			$\mathcal{H}:\{h_{i,j}, \forall i,j \in \mathcal{V}\}$ & Set of number of hops for shortest path from $i$ to $j$.\\
			\rowcolor{Gray}
			$\mathcal{Q}:\{q_i, \forall i \in \mathcal{V}\}$ & Set of switches' processing capacity.\\
			$\mathcal{C}^j \subseteq \mathcal{C}$ & Set of controllers connected to the node $j$.\\
			\rowcolor{Gray}
			$\mathcal{M} \subseteq \mathcal{V}$ & Set of switches connected to at least one controller.\\
			$T$ & Maximum tolerated delay value.\\
			\rowcolor{Gray}
			$x(i,k)$ & Binary variable that equals $1$ if $i$ is a processing node for $k$.\\
			\hline		
		\end{tabular}
		}
		\label{params}
\end{table}

We describe the processing node mapping problem using an integer linear programming (ILP) formulation. Table \ref{params} summarizes the notation used.

\textbf{Communication overhead minimization objective} minimizes the global imposed communication footprint in the control plane. Each controller replica generates an individual message sent to the processing node $i$, that subsequently collects all remaining necessary messages and forwards a resulting single correct message to the configuration target $k$:
\begin{equation}
	\resizebox{\columnwidth}{!}
	{
		$M_F = \text{min\quad}{\sum\limits_{k\in \mathcal{V}}\sum\limits_{i\in \mathcal{V}} (1 \cdot h_{i,k} \cdot x(i,k) +\sum\limits_{j\in \mathcal{M}} |\mathcal{C}^j| \cdot h_{j,i} \cdot x(i,k))}$
		}
		\label{objective1}
\end{equation}
\textbf{Configuration delay minimization objective} minimizes the \emph{worst-case} delay imposed on the critical path used for forwarding configuration messages from a controller associated with node $j$, to the potential processing node $i$ and finally to the configuration target node $k$:
\begin{equation}
	{M_D = \text{min\quad}{\sum\limits_{k\in \mathcal{V}}\sum\limits_{i\in \mathcal{V}} x(i,k) \cdot \underset{\forall j\in \mathcal{M}}{\max}(d_{j,i,k})}}
	\label{objective2}
\end{equation}

\textbf{Bi-objective optimization} minimizes the weighted sum of the two objectives, $w_1$ and $w_2$ being the associated weights: \begin{equation} \text{min\ }w_1 \cdot M_F + w_2 \cdot M_D \label{weightedsum} \end{equation}

	\textbf{Processing capacity constraint:} Sum of messages requiring processing on $i$, for each configuration target $k$ assigned to $i$, must be kept at or below $i$'s processing capacity $q_i$:
	\begin{equation}
		{\text{Subject to:\quad}}{\sum\limits_{k\in \mathcal{V}} x(i,k) \cdot |\mathcal{C}|\leqslant q_i ,\quad \forall i\in \mathcal{V}}
	\end{equation}

	\textbf{Maximum delay constraint:} For each configuration target $k$, the delay imposed by the controller packet forwarding to node $i$, responsible for collection and packet comparison procedure and forwarding of the correct message to the target node $k$, does not exceed an upper bound $T$:
	\begin{align}
		{\text{Subject to:\quad}}{\sum\limits_{i\in \mathcal{V}} x(i,k) \cdot \underset{\forall j\in \mathcal{M}}{\max}(d_{j,i,k})\leqslant T ,\quad \forall k\in \mathcal{V}}
	\end{align}

	\textbf{Single assignment constraint:} For each configuration target $k$, there exists \emph{exactly} one processing node $i$:
	\begin{equation}
		{\text{Subject to:\quad}}{\sum\limits_{i\in \mathcal{V}} x(i,k) = 1 ,\quad \forall k\in \mathcal{V}}
	\end{equation}

	\emph{Note:} The assignment of controller-switch connections for the purpose of control and reconfiguration is adapted from existing formulations \cite{morph, iccbft} and is thus not detailed here.

	\subsection{P4 Switch and Reassigner Control Flow}

	\emph{Processing node data plane:} Switches declared to process controller messages for a particular target (i.e., for itself, or for another switch) initially collect the control payloads stemming from different controllers. Each processing node maintains counters for the number of observed and matching packets for a particular (re-)configuration request identifier. After sufficient matching packets are collected for a particular payload (more specifically, hash of the payload), the processing node \emph{signs} a message using its private key and forwards one copy of the correct packet to its own control plane for required software processing (i.e., identification of the correct message and potentially malicious controllers), and the second copy on the port leading to the configuration target. To distinguish processed from unprocessed packets in destination switches, processing nodes refer to the trailing signature field. 


	\emph{Processing node control plane:} After determining the correct packet, the processing node identifies any incorrect controller replicas (i.e., replicas whose output hashes diverge from the deduced correct hash) and subsequently notifies the Reassigner of the discrepancy. Alternatively, the switch applies the configuration message if it is the configuration target itself. The switch then proceeds to clear its registers associated with the processed message hash so to free the memory for future requests. 

	\emph{Reassigner control flow:} At network bootstrapping time, or on occurrence of any of the following events: i) a detected malicious controller; ii) a failed controller replica; or iii) a switch/link failure; Reassigner reconfigures the processing and forwarding tables of the switches, as well as the number of required matching messages to detect the correct message. 

	\subsection{P4 Tables Design}
	\label{design}
	Switches maintain \emph{Tables} and \emph{Registers} that define the method of processing incoming packets. Reassigner populates the switches' Tables and Registers so that the selection of processing nodes for controller messages is optimal w.r.t. a set of given constraints, i.e., so that the total message overhead or control plane latency experienced in control plane is minimized (according to the optimization procedure in Section \ref{optimization}). The Reassigner thus modifies the elements whenever a controller is identified as incorrect and is hence excluded from consideration, resulting in a different optimization result.  P4BFT leverages four P4 tables:

	\begin{enumerate} 
		\item \emph{Processing Table}: It holds identifiers of the switches whose packets must be processed by the switch hosting this table. Incoming packets are matched based on the destination switch's ID. In the case of a table hit, the hosting switch processes the packets as a processing node. Alternatively, the packet is matched against the \emph{Process-Forwarding Table}. 

		\item \emph{Process-Forwarding Table}: Declares which egress port the packets should be sent out on for further processing. If an unprocessed control packet is not to be processed locally, the switch will forward the packet towards the correct processing node, based on forwarding entries maintained in this table. 

		\item \emph{L2-Forwarding Table}: After the processing node has processed the incoming control packets destined for the destination switch, the last step is forwarding the correctly deduced packet towards it. Information on how to reach the destination switches is maintained in this table. Contrary to forwarding to a processing node, the difference here is that the packet is now forwarded to the destination switch. 



		\item \emph{Hash Table with associated registers}: \emph{Processing} a set of controller packets for a particular request identifier requires evaluating and counting the number of occurrences of packets containing the matching payload. To uniquely identify the decision of the controller, a hash value is generated on the payload during processing. The counting of incoming packets is done by updating the corresponding binary values in the register vectors, with respective layout depicted in Table \ref{hashregisters}.
	\end{enumerate} 


	On each arriving unprocessed packet, the processing node computes a previously seen or $i$-th initially observed hash $h_i^{request\_id}$ over the acquired payload. Subsequently, it sets the binary flag to $1$, for source controller \texttt{controller\_id} in the $i$-th register row at column [$\texttt{client\_request\_id} \cdot |\mathcal{C}|$ + \texttt{controller\_id}]. $|\mathcal{C}|$ represents the total no. of deployed controllers. Each time a client request is fully processed, the binary entries associated with the corresponding \texttt{client\_request\_id} are reset to zero. To detect a malicious controller, the controller IDs associated with hashes distinguished as incorrect, are reported to the Reassigner. 

	\emph{Note:} To tolerate $F_M$ Byzantine failures, a maximum of $F_M+1$ unique hashes for a single request identifier may be detected, hence the corresponding $F_M+1$ pre-allocated table rows in Table \ref{hashregisters}.
	\begin{table}
		\centering
		\caption{Hash Table Layout}	
		\resizebox{\columnwidth}{!}{
			\begin{tabular}{ |M{1.4cm}|M{0.4cm}|M{0.4cm}|M{0.4cm}|M{0.4cm}|M{0.4cm}||M{0.4cm}||M{0.4cm}|M{0.4cm}|M{0.4cm}|M{0.4cm}|M{0.4cm}|}
				\hline
				\textbf{Msg Hash} & \multicolumn{5}{c||}{\textbf{Request ID $1$}} & ... & \multicolumn{5}{c|}{\textbf{Request ID $K$}} \\
				\hline
				$h_0$ & $b_{C_1}^{h_0}$ & $b_{C_2}^{h_0}$ & \multicolumn{2}{c|}{...} & $b_{C_N}^{h_0}$ & ... & $b_{C_1}^{h_0}$ & $b_{C_2}^{h_0}$ & \multicolumn{2}{c|}{...} & $b_{C_N}^{h_0}$ \\
				\hline
				... & $b_{C_1}^{...}$ & $b_{C_2}^{...}$ & \multicolumn{2}{c|}{...} & $b_{C_N}^{...}$ & ... & $b_{C_1}^{...}$ & $b_{C_2}^{...}$ & \multicolumn{2}{c|}{...} & $b_{C_N}^{...}$ \\
				\hline
				$h_{F_M}$ & $b_{C_1}^{h_{F_M}}$ & $b_{C_2}^{h_{F_M}}$ & \multicolumn{2}{c|}{...} & $b_{C_N}^{h_{F_M}}$ & ... & $b_{C_1}^{h_{F_M}}$ & $b_{C_2}^{h_{F_M}}$ & \multicolumn{2}{c|}{...} & $b_{C_N}^{h_{F_M}}$ \\
				\hline
			\end{tabular}
			}
			\label{hashregisters}
	\end{table}

%% file: optimization.tex
\subfloat[Case I: $F_C$ = 13; $F_D$ = 3 hops]{
	\resizebox{.3\textwidth}{!}{%
		\begin{tikzpicture}
			\draw [dashed, thick, orange] (-0.5,3.5) rectangle ++(1.5,1.2);
			\draw [dashed, thick, OliveGreen] (-1,-0.55) rectangle ++(2,1.1);
			\draw [dashed, thick, red] (5.5, 3.5) rectangle ++(1.9,1.2);
			\draw [dashed, thick, red] (5.5, 3.5) rectangle ++(1.9,1.2);
			\node (label1) at (0.25, 4.5) {Cluster 1};
			\node (label2) at (6.5, 4.5) {Cluster 2};
			\node (C1) at (0,4)[circle, draw=black, fill=white, inner sep=0pt, minimum size = 15pt] {C1};
			\node (C2) at (0.5,4)[circle, draw=black, fill=white, inner sep=0pt, minimum size = 15pt] {C2};
			\node (S1) at (0,2){\includegraphics{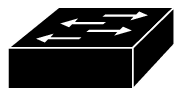}}; 
			\node (S2) at (3,2) {\includegraphics{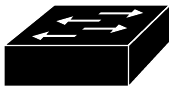}}; 
			\node (S3) at (6,2) {\includegraphics{figures/assets/switch}};
			\node (S4) at (0,0) {\includegraphics{figures/assets/switch}};
			\node (S5) at (6,0) {\includegraphics{figures/assets/switch}};
			\node (C3) at (6,4) [circle, draw=black, fill=white, inner sep=0pt, minimum size = 15pt] {C3};
			\node (C4) at (6.5,4) [circle, draw=black, fill=white, inner sep=0pt, minimum size = 15pt] {C4};
			\node (C5) at (7,4) [circle, draw=black, fill=white, inner sep=0pt, minimum size = 15pt] {C5};
			\draw(S1) -- (S2);
			\draw(S2) -- (S3);
			\draw(S1) -- (S4);
			\draw(S4) -- (S5);
			\draw(S4) -- (S2);
			\draw(S3) -- (S5);
			\draw(C1) -- (S1);
			\draw(C2) -- (S1);
			\draw(S3) -- (C3);
			\draw(S3) -- (C4);
			\draw(S3) -- (C5);

			\draw[thick, orange, ->, -Latex]([xshift=0.1cm]C1.south) -- ([xshift=0.1cm]S1.north);
			\draw[thick, orange, ->, -Latex]([xshift=0.1cm]C2.south) -- ([xshift=0.3cm]S1.north);
			\draw[thick, orange, ->, -Latex]([xshift=0.1cm]S1.south) -- ([xshift=0.1cm]S4.north);
			\draw[thick, orange, ->, -Latex]([xshift=0.2cm]S1.south) -- ([xshift=0.2cm]S4.north);
			\draw[thick, red, ->, -Latex]([xshift=0.1cm]C3.south) -- ([xshift=0.1cm]S3.north);
			\draw[thick, red, ->, -Latex]([xshift=0.1cm]C4.south) -- ([xshift=0.2cm]S3.north);
			\draw[thick, red, ->, -Latex]([xshift=0.1cm]C5.south) -- ([xshift=0.3cm]S3.north);
			\draw[thick, red, ->, -Latex]([yshift=0.1cm]S3.west) -- ([yshift=0.1cm]S2.east);
			\draw[thick, red, ->, -Latex]([yshift=0.2cm]S3.west) -- ([yshift=0.2cm]S2.east);
			\draw[thick, red, ->, -Latex]([yshift=0.0cm, xshift=-0.05cm]S2.south west) -- ([xshift=-0.3cm]S4.north east);
			\draw[thick, red, ->, -Latex]([yshift=0.0cm, xshift=0.3cm]S2.south west) -- ([xshift=0.03cm]S4.north east);
			\draw[thick, red, ->, -Latex]([xshift=0.1cm]S3.south) -- ([xshift=0.1cm]S5.north);
			\draw[thick, red, ->, -Latex]([yshift=0.1cm]S5.west) -- ([yshift=0.1cm]S4.east);

			\node at ([shift=({0.8cm,1cm})]S1)[orange] {+2};
			\node at ([shift=({0.5cm,1cm})]S4)[orange] {+2};
			\node at ([shift=({2cm,1cm})]S4)[red] {+2};
			\node at ([shift=({2cm,0.7cm})]S4)[blue] {+1h};
			\node at ([shift=({0.5cm,1cm})]S5)[red] {+1};
			\node at ([shift=({-3cm,0.4cm})]S5)[red] {+1};
			\node at ([shift=({-1.5cm,0.5cm})]S3)[blue] {+1h};
			\node at ([shift=({-1.5cm,0.9cm})]S3)[red] {+2};
			\node at ([shift=({1cm,1cm})]S3)[red] {+3};
			\node at ([shift=({1cm,0.6cm})]S3)[blue] {+1h};
			\node at ([shift=({-0.3cm,-0.2cm})]S1)[white] {S1};
			\node at ([shift=({-0.3cm,-0.2cm})]S2)[white] {S2};
			\node at ([shift=({-0.3cm,-0.2cm})]S3)[white] {S3};
			\node at ([shift=({-0.3cm,-0.2cm})]S4)[white] {S4};
			\node at ([shift=({-0.3cm,-0.2cm})]S5)[white] {S5};
		\end{tikzpicture}
	}
}
\hfill
\subfloat[Case II: $F_C$ = 11; $F_D$ = 5 hops]{
	\resizebox{.3\textwidth}{!}{%
		\begin{tikzpicture}
			\draw [dashed, thick, orange] (-0.5,3.5) rectangle ++(1.5,1.2);
			\draw [dashed, thick, red] (5.5, 3.5) rectangle ++(1.9,1.2);
			\draw [dashed, thick, OliveGreen] (5.05,1.45) rectangle ++(2,1.1);
			\node (label1) at (0.25, 4.5) {Cluster 1};
			\node (label2) at (6.5, 4.5) {Cluster 2};
			\node (C1) at (0,4)[circle, draw=black, fill=white, inner sep=0pt, minimum size = 15pt] {C1};
			\node (C2) at (0.5,4)[circle, draw=black, fill=white, inner sep=0pt, minimum size = 15pt] {C2};
			\node (S1) at (0,2){\includegraphics{figures/assets/switch}}; 
			\node (S2) at (3,2) {\includegraphics{figures/assets/switch}}; 
			\node (S3) at (6,2) {\includegraphics{figures/assets/switch}};
			\node (S4) at (0,0) {\includegraphics{figures/assets/switch}};
			\node (S5) at (6,0) {\includegraphics{figures/assets/switch}};
			\node (C3) at (6,4) [circle, draw=black, fill=white, inner sep=0pt, minimum size = 15pt] {C3};
			\node (C4) at (6.5,4) [circle, draw=black, fill=white, inner sep=0pt, minimum size = 15pt] {C4};
			\node (C5) at (7,4) [circle, draw=black, fill=white, inner sep=0pt, minimum size = 15pt] {C5};
			\draw(S1) -- (S2);
			\draw(S2) -- (S3);
			\draw(S1) -- (S4);
			\draw(S4) -- (S5);
			\draw(S4) -- (S2);
			\draw(S3) -- (S5);
			\draw(C1) -- (S1);
			\draw(C2) -- (S1);
			\draw(S3) -- (C3);
			\draw(S3) -- (C4);
			\draw(S3) -- (C5);

			\draw[thick, orange, ->, -Latex]([xshift=0.1cm]C1.south) -- ([xshift=0.1cm]S1.north);
			\draw[thick, orange, ->, -Latex]([xshift=0.1cm]C2.south) -- ([xshift=0.3cm]S1.north);
			\draw[thick, orange, ->, -Latex]([yshift=0.1cm]S1.east) -- ([yshift=0.1cm]S2.west);
			\draw[thick, orange, ->, -Latex]([yshift=-0.1cm]S1.east) -- ([yshift=-0.1cm]S2.west);
			\draw[thick, orange, ->, -Latex]([yshift=0.1cm]S2.east) -- ([yshift=0.1cm]S3.west);
			\draw[thick, orange, ->, -Latex]([yshift=-0.1cm]S2.east) -- ([yshift=-0.1cm]S3.west);

			\draw[thick, red, ->, -Latex]([xshift=0.1cm]C3.south) -- ([xshift=0.1cm]S3.north);
			\draw[thick, red, ->, -Latex]([xshift=0.1cm]C4.south) -- ([xshift=0.2cm]S3.north);
			\draw[thick, red, ->, -Latex]([xshift=0.1cm]C5.south) -- ([xshift=0.3cm]S3.north);

			\draw[thick, OliveGreen, ->, -Latex]([yshift=0.2cm]S3.west) -- ([yshift=0.2cm]S2.east);
			\draw[thick, OliveGreen, ->, -Latex]([yshift=0.0cm, xshift=-0.05cm]S2.south west) -- ([xshift=-0.3cm]S4.north east);

			\node at ([shift=({0.8cm,1.1cm})]S1)[orange] {+2};
			\node at ([shift=({0.8cm,0.8cm})]S1)[blue] {+1h};
			\node at ([shift=({1.5cm,0.7cm})]S1)[orange] {+2};
			\node at ([shift=({1.5cm,0.4cm})]S1)[blue] {+1h};
			\node at ([shift=({2cm,1.1cm})]S4)[OliveGreen] {+1};
			\node at ([shift=({2cm,0.8cm})]S4)[blue] {+1h};
			\node at ([shift=({-1.5cm,0.8cm})]S3)[OliveGreen] {+1};
			\node at ([shift=({-1.5cm,0.5cm})]S3)[blue] {+1h};
			\node at ([shift=({-1.5cm,-0.5cm})]S3)[orange] {+2};
			\node at ([shift=({-1.5cm,-0.8cm})]S3)[blue] {+1h};
			\node at ([shift=({1cm,1cm})]S3)[red] {+3};

			\node at ([shift=({-0.3cm,-0.2cm})]S1)[white] {S1};
			\node at ([shift=({-0.3cm,-0.2cm})]S2)[white] {S2};
			\node at ([shift=({-0.3cm,-0.2cm})]S3)[white] {S3};
			\node at ([shift=({-0.3cm,-0.2cm})]S4)[white] {S4};
			\node at ([shift=({-0.3cm,-0.2cm})]S5)[white] {S5};
		\end{tikzpicture}
	}
}
\hfill
\subfloat[Case III: $F_C$ = 11; $F_D$ = 3 hops]{
	\resizebox{.3\textwidth}{!}{%
		\begin{tikzpicture}
			\draw [dashed, thick, orange] (-0.5,3.5) rectangle ++(1.5,1.2);
			\draw [dashed, thick, red] (5.5, 3.5) rectangle ++(1.9,1.2);
			\draw [dashed, thick, OliveGreen] (2,1.45) rectangle ++(2,1.1);
			\node (label1) at (0.25, 4.5) {Cluster 1};
			\node (label2) at (6.5, 4.5) {Cluster 2};
			\node (C1) at (0,4)[circle, draw=black, fill=white, inner sep=0pt, minimum size = 15pt] {C1};
			\node (C2) at (0.5,4)[circle, draw=black, fill=white, inner sep=0pt, minimum size = 15pt] {C2};
			\node (S1) at (0,2){\includegraphics{figures/assets/switch}}; 
			\node (S2) at (3,2) {\includegraphics{figures/assets/switch}}; 
			\node (S3) at (6,2) {\includegraphics{figures/assets/switch}};
			\node (S4) at (0,0) {\includegraphics{figures/assets/switch}};
			\node (S5) at (6,0) {\includegraphics{figures/assets/switch}};
			\node (C3) at (6,4) [circle, draw=black, fill=white, inner sep=0pt, minimum size = 15pt] {C3};
			\node (C4) at (6.5,4) [circle, draw=black, fill=white, inner sep=0pt, minimum size = 15pt] {C4};
			\node (C5) at (7,4) [circle, draw=black, fill=white, inner sep=0pt, minimum size = 15pt] {C5};
			\draw(S1) -- (S2);
			\draw(S2) -- (S3);
			\draw(S1) -- (S4);
			\draw(S4) -- (S5);
			\draw(S4) -- (S2);
			\draw(S3) -- (S5);
			\draw(C1) -- (S1);
			\draw(C2) -- (S1);
			\draw(S3) -- (C3);
			\draw(S3) -- (C4);
			\draw(S3) -- (C5);

			\draw[thick, orange, ->, -Latex]([xshift=0.1cm]C1.south) -- ([xshift=0.1cm]S1.north);
			\draw[thick, orange, ->, -Latex]([xshift=0.1cm]C2.south) -- ([xshift=0.3cm]S1.north);
			\draw[thick, orange, ->, -Latex]([yshift=0.1cm]S1.east) -- ([yshift=0.1cm]S2.west);
			\draw[thick, orange, ->, -Latex]([yshift=-0.1cm]S1.east) -- ([yshift=-0.1cm]S2.west);

			\draw[thick, red, ->, -Latex]([xshift=0.1cm]C3.south) -- ([xshift=0.1cm]S3.north);
			\draw[thick, red, ->, -Latex]([xshift=0.1cm]C4.south) -- ([xshift=0.2cm]S3.north);
			\draw[thick, red, ->, -Latex]([xshift=0.1cm]C5.south) -- ([xshift=0.3cm]S3.north);

			\draw[thick, red, ->, -Latex]([yshift=0.1cm]S3.west) -- ([yshift=0.1cm]S2.east);
			\draw[thick, red, ->, -Latex]([yshift=0.2cm]S3.west) -- ([yshift=0.2cm]S2.east);
			\draw[thick, red, ->, -Latex]([yshift=-0.1cm]S3.west) -- ([yshift=-0.1cm]S2.east);

			\draw[thick, OliveGreen, ->, -Latex]([yshift=0.0cm, xshift=-0.05cm]S2.south west) -- ([xshift=-0.3cm]S4.north east);

			\node at ([shift=({0.8cm,1cm})]S1)[orange] {+2};
			\node at ([shift=({1.5cm,0.5cm})]S1)[orange] {+2};
			\node at ([shift=({2cm,1cm})]S4)[OliveGreen] {+1};
			\node at ([shift=({2cm,0.7cm})]S4)[blue] {+1h};
			\node at ([shift=({-1.5cm,0.8cm})]S3)[red] {+3};
			\node at ([shift=({-1.5cm,0.5cm})]S3)[blue] {+1h};
			\node at ([shift=({1cm,1cm})]S3)[red] {+3};
			\node at ([shift=({1cm,0.7cm})]S3)[blue] {+1h};

			\node at ([shift=({-0.3cm,-0.2cm})]S1)[white] {S1};
			\node at ([shift=({-0.3cm,-0.2cm})]S2)[white] {S2};
			\node at ([shift=({-0.3cm,-0.2cm})]S3)[white] {S3};
			\node at ([shift=({-0.3cm,-0.2cm})]S4)[white] {S4};
			\node at ([shift=({-0.3cm,-0.2cm})]S5)[white] {S5};
		\end{tikzpicture}
	}
}

%% file: discussion.tex
\section{Evaluation, Results and Discussion}
\label{discussion}

\subsection{Evaluation Methodology}

We next evaluate the following metrics using P4BFT and state-of-the-art \cite{morph, mohan2017primary, li2014byzantine} designs: i) control plane load; ii) imposed processing delay in the software and hardware P4BFT nodes; iii) end-to-end switch reconfiguration delay; and iv) ILP solution time. We execute the measurements for random controller placements and diverse data plane topologies: i) random topologies with fixed average node degree; ii) reference Internet2 \cite{internet2}; and iii) data-center Fat-Tree ($k=4$). We also vary and depict the impact of no. of switches, controller instances, and disjoint controller clusters. To compute paths between controllers and switches and between processing and destination switches, Reassigner leverages the Constrained Shortest Path First (CSPF) algorithm. For brevity, as an input to the optimization procedure in Reassigner, we assume edge weights of $1$. The objective function used in processing node selection is Eq. \ref{weightedsum}, parametrized with $(w_1, w_2) = (1,1)$.

P4BFT implementation is a combination of $\text{P4}_{\text{16}}$ and P4 Runtime code, compiled for software and physical execution on P4 software switch \emph{bmv2} (\emph{master} check-out, December 2018) and a \emph{Netronome Agilio SmartNIC} device with the corresponding firmware compiled using \texttt{SDK 6.1-Preview}, respectively. Apache Thrift and gRPC are used for population of registers and table entries in \emph{bmv2}, respectively. Thrift is used for both table and registers population for the \emph{Netronome SmartNIC}, due to the current SDK release not fully supporting the P4 Runtime. HTTP REST is used in exchange between P4 switch control plane and the Reassigner. The Reassigner and network controller replicas are implemented as Python applications.

\subsection{Communication Overhead Advantage}
\begin{figure}[htb]
	\centering
	\includegraphics[width=.45\textwidth]{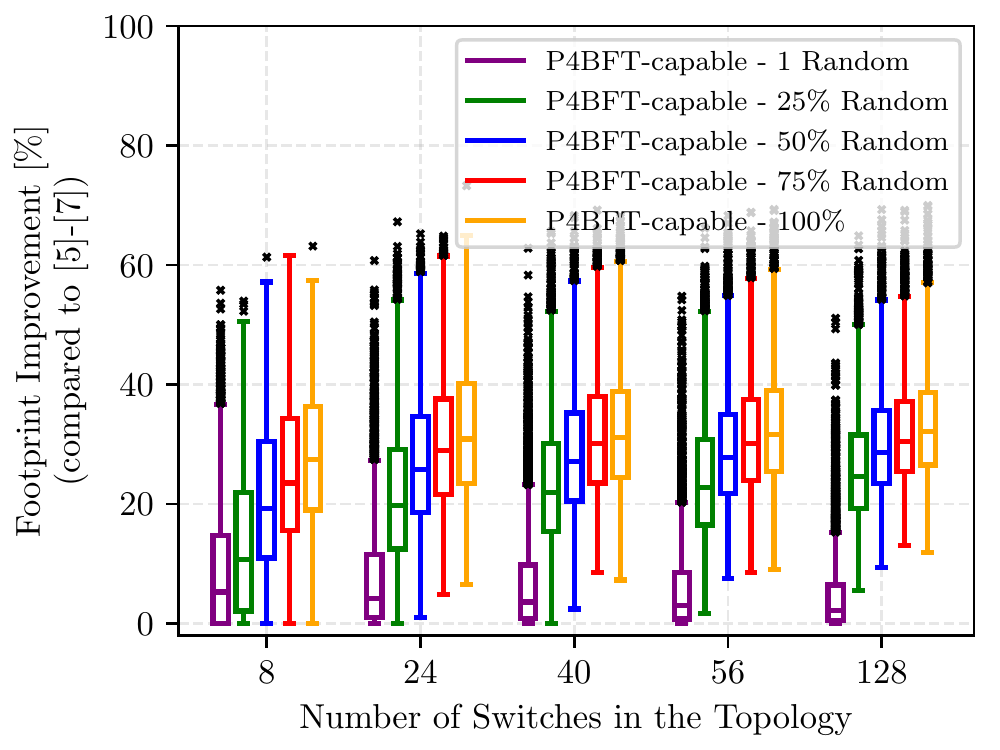}
	\caption{Packet load improvement of P4BFT over the reference works \cite{morph, li2014byzantine, mohan2017primary} for $5000$ randomly generated network topologies per scenario, with $7$ controllers distributed into $3$ disjoint and randomly placed clusters. In addition to the 100\% coverage where each node may be considered a P4BFT processing node, we include scenarios where only the random [1, 25\%, 50\%, 75\%] nodes of all available nodes in the infrastructure are P4BFT-enabled. Thus, even in the topologies with limited programmable data plane resources, i.e., in brownfield-scenarios involving OpenFlow/NETCONF+YANG non-P4 configuration targets, P4BFT offers substantial advantages over existing SoA.} 
	\label{fig:footprint_switches}
\end{figure}

Figure \ref{fig:footprint_switches} depicts the packet load improvement in P4BFT over the existing reference solutions \cite{morph, li2014byzantine, mohan2017primary} for randomly generated topologies with average node degree of $4$. The footprint improvement is defined as $1-\frac{F_C^{P4BFT}}{F_C^{SoA}}$, where $F_C$ denotes the sum of packet footprint for control flows destined to each destination switch of the network topology as per Sec. \ref{optimization} and Fig. \ref{fig:failure_model}. P4BFT outperforms the state-of-the-art as each of the presented works assumes an uninterrupted control flow from each controller instance to the destination switches. P4BFT, on the other hand, aggregates control packets in the processing nodes that, subsequently to collecting the control packets, forward a single correct message towards the destination, thus decreasing the control plane load. 

\begin{figure}[htb]
	\centering
	\subfloat[]{\includegraphics[width=.4\textwidth]{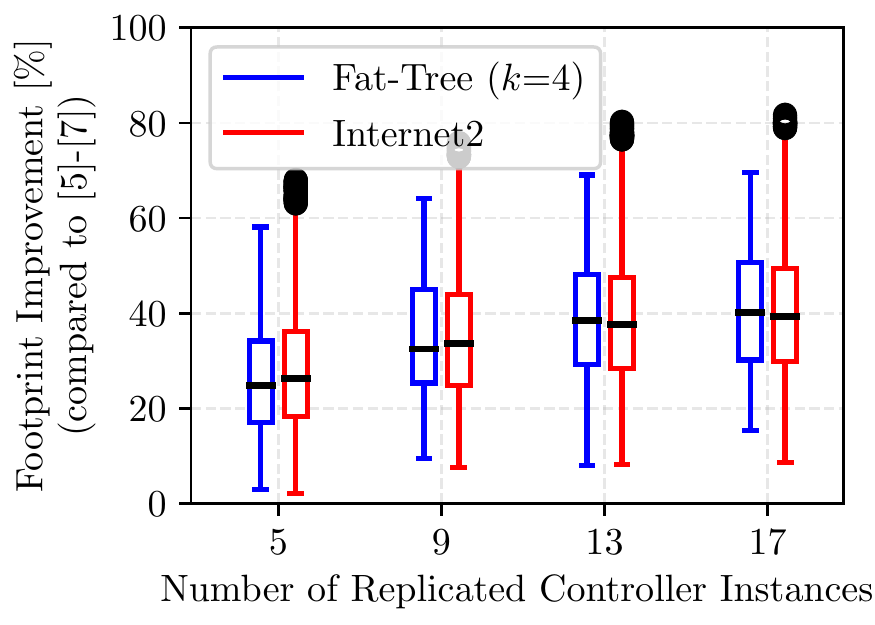}}

	\subfloat[]{\includegraphics[width=.4\textwidth]{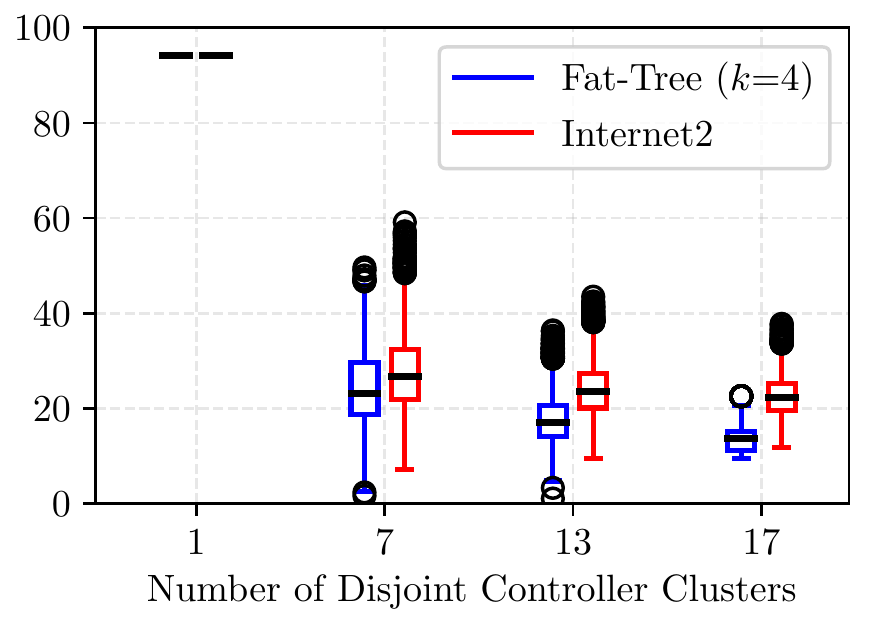}}
	\caption{The impact of (a) controllers and; (b) disjoint controller clusters on the control plane load footprint in Internet2 and Fat-Tree ($k=4$) topologies for $5000$ randomized controller placements each. (a) randomizes the placement but fixes the no. of disjoint clusters to $3$; (b) randomizes the no. of disjoint clusters between [1, 7, 13, 17] but fixes the no. of controllers to $17$. The resulting footprint improvement scales with the number of controllers but is inversely proportional to the number of disjoint clusters.}
\label{fig:footprint_controllers}
\end{figure}

Fig. \ref{fig:footprint_controllers} (a) and (b) portray the footprint improvement scaling with the number of controllers and disjoint clusters. P4BFT's footprint efficiency generally benefits from the higher number of controller instances. Controller clusters, on the other hand, aggregate replicas behind the same edge switch. Thus, with the higher number of disjoint clusters, the degree of aggregation and the total footprint improvement decreases. 

\subsection{Processing and Reconfiguration Delay}

\begin{figure}[htb]
	\centering
	\includegraphics[width=.48\textwidth]{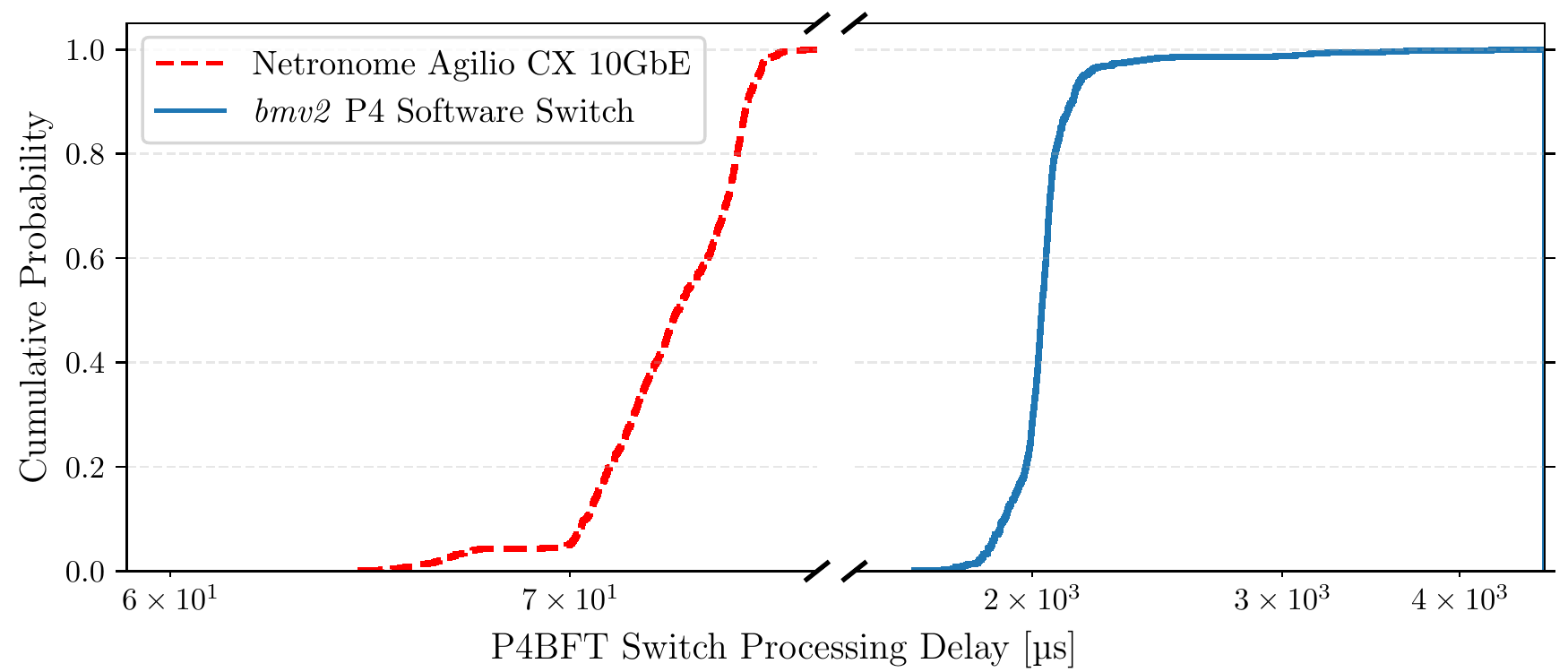}
	\caption{The CDF of processing delays imposed in a P4BFT's processing node for a scenario including $5$ controller instances. $3$ correct packets and thus $3$ P4 pipeline executions are necessary to confirm the payload correctness when tolerating $2$ Byzantine controller failures.}
	\label{fig:processing_delay}
\end{figure}

Fig. \ref{fig:processing_delay} depicts the processing delay incurred in the processing node for a single client request. The delay corresponds to the P4 pipeline execution time spent on identification of a correct controller message, comprising the i) hash computation over controller messages; ii) incrementing the counters for the computed hash; iii) signing the correct packet and; iv) propagating it to the correct egress port. When using the P4-enabled SmartNIC, P4BFT decreases the processing time compared to \emph{bmv2} software target by two orders of magnitude.

\begin{figure}[htb]
	\centering
	\includegraphics[width=.48\textwidth]{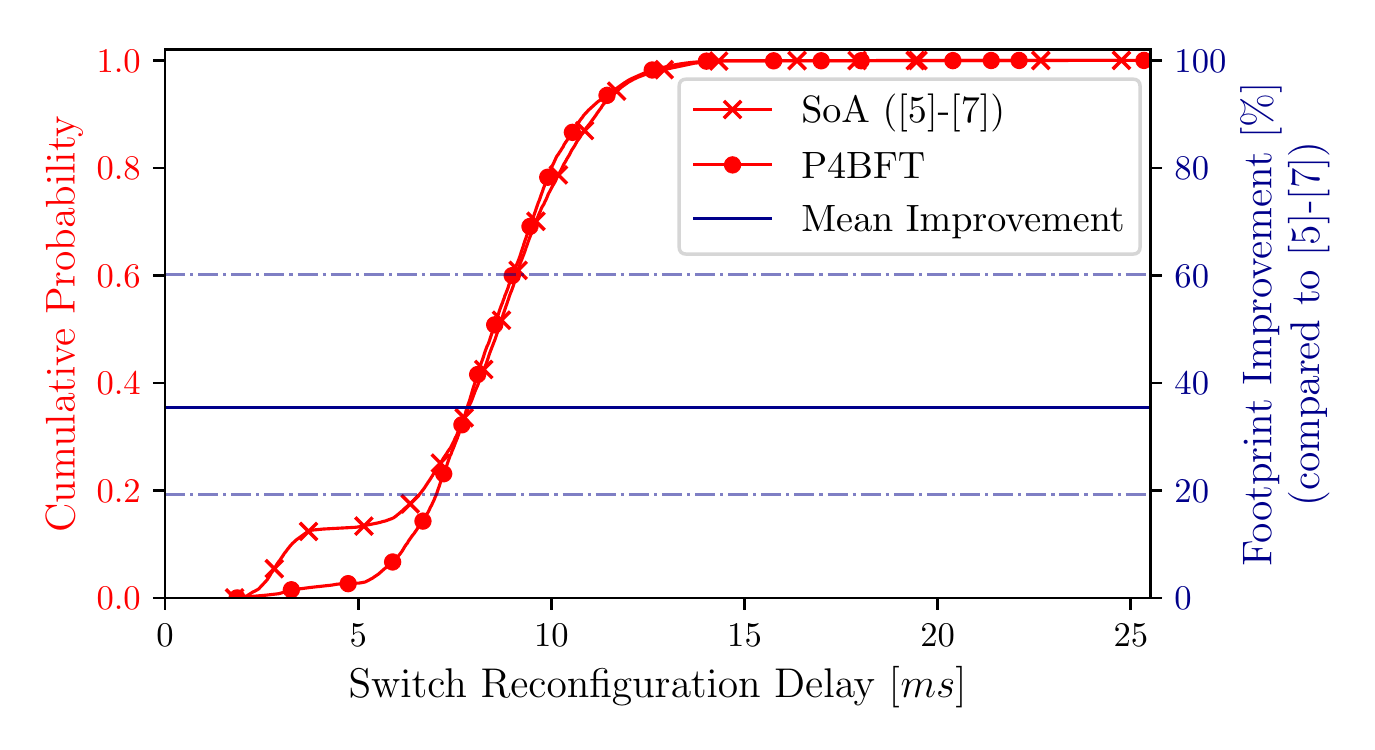}
	\caption{CDFs of time taken to configure randomly selected switches in SoA and P4BFT environments for Internet2 topology, $10$ random controller placements for $5$ replicas and $1700$ individual requests per placement. SoA works [5]-[7] collect, compare and apply the controllers' reconfiguration messages in the destination switch thus effectively minimizing the reconfiguration delay at all times. P4BFT, on the other hand, may occasionally favor footprint minimization over the incurred reconfiguration delay and thus impose a longer critical path, leading to slower reconfigurations. On average, however, P4BFT imposes comparable reconfiguration delays at a much higher footprint improvement (depicted blue), mean being $38\%$, best and worst cases at $60\%$ and $19.3\%$, respectively, for evaluated placements.} 
	\label{fig:reconfig_delay}
\end{figure}

Fig. \ref{fig:reconfig_delay} depicts the total reconfiguration delay imposed in SoA and P4BFT designs for $(w_1, w_2) = (1,1)$ (ref. Eq. \ref{weightedsum}). It considers the time difference between issuing a switch reconfiguration request, until the correct controller message is determined and applied in the destination. Related works process the reconfiguration messages stemming from controller replicas in the destination target, their control flows traversing shortest paths in all cases. On average, P4BFT's reconfiguration delay is comparable with related works, the overall control plane footprint being substantially improved.

\subsection{Optimization procedure in Reassigner}
\subsubsection{Impact of optimization objectives} 
\begin{figure}[htb]
	\centering
	\includegraphics[width=.45\textwidth]{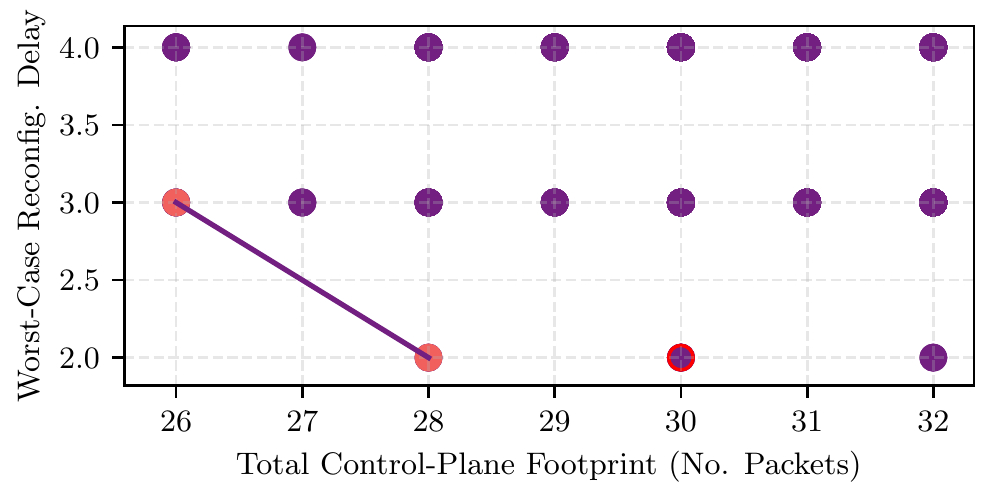}
	\caption{Pareto Frontier of P4BFT's solution space for the topology presented in Fig. \ref{fig:failure_model}. The comparable works tend to minimize the incurred reconfiguration delay, but ignore the imposed control plane load. \cite{morph, li2014byzantine, mohan2017primary} hence select $(30.0, 2.0)$ as the optimal solution (encircled in red) while P4BFT selects $(26.0, 3.0)$ or $(28.0, 2.0)$ thus minimizing the total overhead as per Eq. \ref{weightedsum}.}
	\label{fig:pareto}
\end{figure}

Figure \ref{fig:pareto} depicts the Pareto frontier of optimal processing node assignments w.r.t. the objectives presented in Section \ref{optimization}: the total control plane footprint (minimized as per Eq. \ref{objective1}) and the reconfiguration delay (minimized as per Eq. \ref{objective2}). From the total solution space, depending on the weights prioritization in Eq. \ref{weightedsum}, either $(26.0, 3.0)$ or $(28.0, 2.0)$ solutions can be considered optimal. Comparable works implicitly minimize the incurred reconfiguration delay but fail to consider the control plane load. Hence, they prefer the $(30.0, 2.0)$ solution (encircled red).

\begin{figure}[htb]
	\centering
	\subfloat[]{\includegraphics[width=.45\textwidth]{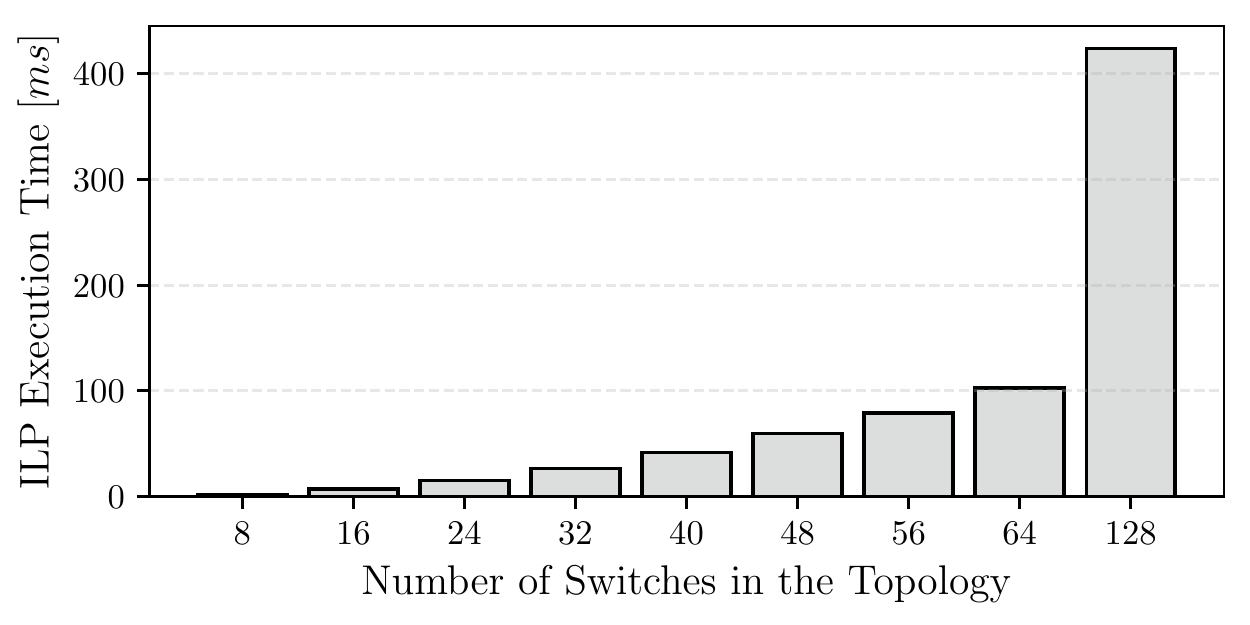}}

	\subfloat[]{\includegraphics[width=.45\textwidth]{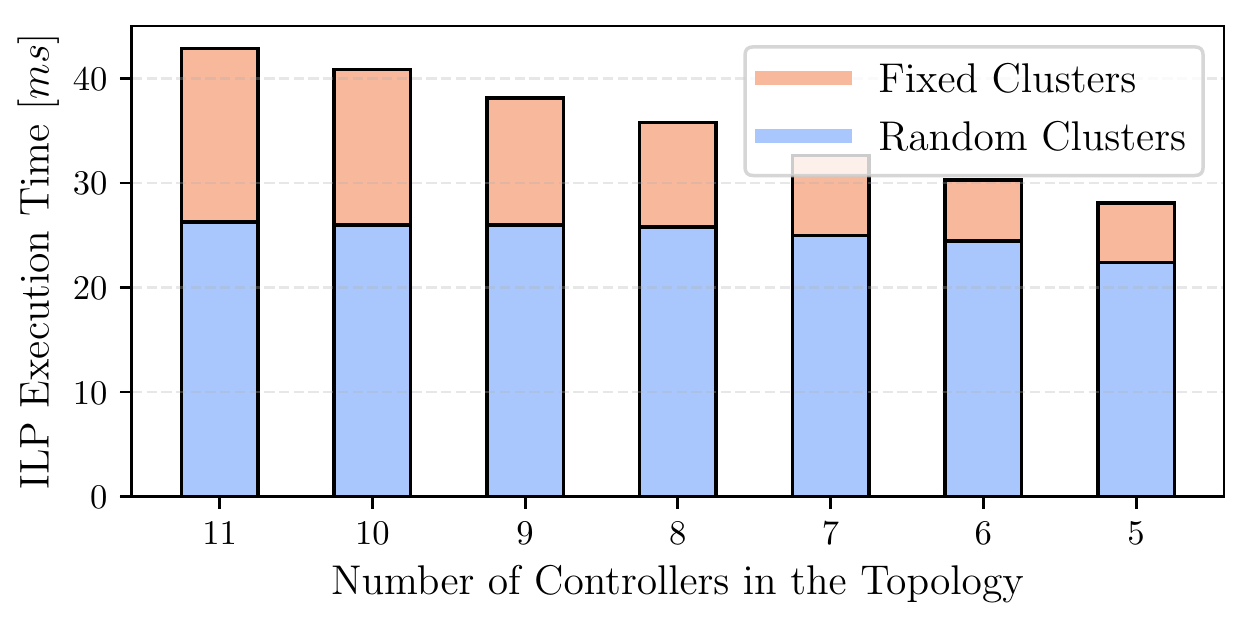}}
	\caption{(a) depicts the impact of network topology size on the ILP solution time for random topologies. (b) depicts the impact of controller number and cluster disjointness in the case of Internet2 topology. The results are averaged over $5000$ per-scenario iterations. The higher the cluster aggregation of controllers, the lower the ILP solution time. "Fixed Clusters" considers the worst-case, where each controller is randomly placed, but disjointly w.r.t. the other controller instances. Clearly, the ILP solution time scales with the amount of deployed switches and controllers. We used Gurobi 8.1 optimization framework configured to execute multiple solvers on multiple threads simultaneously, and have chosen the ones that finish first.}
	\label{fig:ilp}
\end{figure}

\subsubsection{ILP solution time - impact of topology, amount of controller and disjoint clusters} The solution time for the optimization procedure considering random topologies with average network degree of $4$ and a fixed no. of randomly placed controllers is depicted in Fig. \ref{fig:ilp} (a). The solution time scales with number of switches, peaking at $420ms$ for large $128$-switch topologies. The reassignment procedure is executed in few rare events: during network bootstrapping, on malicious / failed controller detection and following a switch / link failure. Thus, we consider the observed solution time short and viable for online mapping. Fig. \ref{fig:ilp} (b) depicts the ILP solution time scaling with the number of active controllers. The lower the number of active controllers, the shorter the solution time. In "Fixed Clusters" case, each controller is placed in its disjoint cluster (worst-case for the optimization). The "Random Clusters" case considers a typical clustering scenario, where a maximum of $[1..3]$ clusters are deployed, each comprising a uniform number of controller instances. The higher the cluster aggregation, the lower the ILP solution time.


%% file: conclusion.tex
\section{Conclusion}
\label{conclusion}

P4BFT introduces a switch control-plane/data-plane co-design, capable of malicious controller identification while simultaneously minimizing the control plane footprint. 
By merging the control channels in P4-enabled \emph{processing} nodes, the use of P4BFT results in a lowered control plane footprint, compared to existing designs. In a hardware-based data plane, by offloading packet processing from general purpose CPU to the data-plane NPU, it additionally leads to a decrease in request processing time. 
Given the low solution time, the presented ILP formulation is viable for on-line execution. While we focused on an SDN scenario here, future works should consider the conceptual transfer of P4BFT to other application domains, including stateful web applications and critical industrial control systems.